\documentclass[preprint, superscriptaddress, aps]{revtex4-2}

\usepackage[english]{babel}
\usepackage{
amsmath, 
amssymb, bbm, mathrsfs, bm, braket, color, graphicx, comment, amsfonts, dsfont}
\usepackage[colorlinks, linkcolor=blue, citecolor=blue, urlcolor=blue]{hyperref}
\usepackage{hyperref}
\usepackage[mathscr]{euscript}
\usepackage{bm}

\usepackage{titlesec}

\begin{document}

\title{Topological nodal $i$-wave superconductivity in PtBi$_2$}

\author{Susmita Changdar}
\affiliation{Leibniz Institute for Solid State and Materials Research, IFW Dresden, Helmholtzstraße 20, 01069 Dresden, Germany}
\affiliation{Institute for Solid State and Materials Physics, TU Dresden, D-01062 Dresden, Germany}
\affiliation{Department of Condensed Matter and Materials Physics, S. N. Bose National Centre for Basic Sciences, Kolkata, West Bengal 700106, India}

\author{Oleksandr Suvorov}
\affiliation{Leibniz Institute for Solid State and Materials Research, IFW Dresden, Helmholtzstraße 20, 01069 Dresden, Germany}
\affiliation{Kyiv Academic University, 36 Vernadsky blvd, Kyiv, 03142, Ukraine}

\author{Andrii Kuibarov}
\affiliation{Leibniz Institute for Solid State and Materials Research, IFW Dresden, Helmholtzstraße 20, 01069 Dresden, Germany}

\author{Setti Thirupathaiah}
\affiliation{Department of Condensed Matter and Materials Physics, S. N. Bose National Centre for Basic Sciences, Kolkata, West Bengal 700106, India}

\author{Grigoriy Shipunov}
\affiliation{Leibniz Institute for Solid State and Materials Research, IFW Dresden, Helmholtzstraße 20, 01069 Dresden, Germany}

\author{Saicharan Aswartham}
\affiliation{Leibniz Institute for Solid State and Materials Research, IFW Dresden, Helmholtzstraße 20, 01069 Dresden, Germany}

\author{Sabine Wurmehl}
\affiliation{Leibniz Institute for Solid State and Materials Research, IFW Dresden, Helmholtzstraße 20, 01069 Dresden, Germany}

\author{Iryna Kovalchuk}
\affiliation{Leibniz Institute for Solid State and Materials Research, IFW Dresden, Helmholtzstraße 20, 01069 Dresden, Germany}
\affiliation{Kyiv Academic University, 36 Vernadsky blvd, Kyiv, 03142, Ukraine}

\author{Klaus Koepernik}
\affiliation{Leibniz Institute for Solid State and Materials Research, IFW Dresden, Helmholtzstraße 20, 01069 Dresden, Germany}

\author{Carsten Timm}
\affiliation{Institute of Theoretical Physics, TU Dresden, 01062 Dresden, Germany}
\affiliation{Würzburg–Dresden Cluster of Excellence ct.qmat, TU Dresden, 01062 Dresden, Germany}

\author{Bernd B\"uchner}
\affiliation{Leibniz Institute for Solid State and Materials Research, IFW Dresden, Helmholtzstraße 20, 01069 Dresden, Germany}
\affiliation{Würzburg–Dresden Cluster of Excellence ct.qmat, TU Dresden, 01062 Dresden, Germany}

\author{Ion Cosma Fulga}
\affiliation{Leibniz Institute for Solid State and Materials Research, IFW Dresden, Helmholtzstraße 20, 01069 Dresden, Germany}
\affiliation{Würzburg–Dresden Cluster of Excellence ct.qmat, TU Dresden, 01062 Dresden, Germany}

\author{Sergey Borisenko}
\affiliation{Leibniz Institute for Solid State and Materials Research, IFW Dresden, Helmholtzstraße 20, 01069 Dresden, Germany}
\affiliation{Würzburg–Dresden Cluster of Excellence ct.qmat, TU Dresden, 01062 Dresden, Germany}

\author{Jeroen van den Brink}
\affiliation{Leibniz Institute for Solid State and Materials Research, IFW Dresden, Helmholtzstraße 20, 01069 Dresden, Germany}
\affiliation{Würzburg–Dresden Cluster of Excellence ct.qmat, TU Dresden, 01062 Dresden, Germany}

\date{\today}

\maketitle

\textbf{
Most superconducting materials are well-understood and {\it conventional} in the sense that the pairs of electrons that cause the superconductivity by their condensation have the highest possible symmetry.
Famous exceptions are the enigmatic high-$T_c$ cuprate superconductors \cite{Sigrist1991}. 
Nodes in their superconducting gap are the fingerprint of their {\it unconventional} character and imply superconducting pairing of $d$-wave symmetry.
Here, using angle-resolved photoemission spectroscopy, we observe that the Weyl semimetal PtBi$_2$ harbors nodes in its superconducting gap, implying unconventional $i$-wave pairing symmetry.
At temperatures below $10\,\mathrm{K}$, the superconductivity in PtBi$_2$ gaps out its topological surface states, the Fermi arcs, while its bulk states remain normal \cite{Kuibarov2024}.
The nodes in the superconducting gap that we observe are located exactly at the center of the Fermi arcs, and imply the presence of topologically protected Majorana cones around this locus in momentum space. 
From this, we infer theoretically that robust zero-energy Majorana flat bands emerge at surface step edges.
This not only establishes PtBi$_2$ surfaces as unconventional, topological $i$-wave superconductors but also as a promising material platform in the ongoing effort to generate and manipulate Majorana bound states.
} 

Electrons in conventional, textbook superconductors such as lead or niobium form Cooper pairs with zero angular momentum ($l=0$) and their pairing symmetry is referred to as $s$-wave.   
Pairing with higher angular momentum and unconventional superconductivity has been established in cuprate high temperature superconductors such as YBa$_2$Cu$_3$O$_7$ and Bi$_2$Sr$_2$CaCu$_2$O$_{8+x}$. 
Their $d$-wave pairing ($l=2$) implies the existence of nodes in the superconducting (SC) gap, locations in momentum space on the Fermi surface where the SC gap vanishes. 
To establish the presence of such nodes in $d$-wave cuprates, angle resolved photoemission spectroscopy (ARPES) has played a pivotal role as it can directly map out the size of the SC gap in momentum space \cite{Shen1993, Ding1995, Damascelli2003, Hashimoto2014, Sobota2021}.

While there is substantial theoretical work discussing SC states with pairing symmetry beyond $l=2$, so far there is no spectroscopic evidence for unconventional superconductivity beyond $d$-wave \cite{Sauls1994, Ikeda2004, Mao2011, Stewart2017}. 
This makes our ARPES-based observation of nodal superconductivity on the Fermi arcs of PtBi$_2$ stand out since a symmetry analysis of its nodal structure implies that the gap here exhibits $i$-wave symmetry ($l=6$).
Since Fermi-arc states are chiral and nondegenerate this sign change in the SC order parameter along the arc implies the formation of a surface Majorana cone, similar to the Majorana cones expected to occur on the surface of three-dimensional (3D) strong topological superconductors \cite{Schnyder2008} or $^3$He \cite{Salomaa1988}, rendering PtBi$_2$ a \textit{topological} superconductor. 
This is remarkable since materials with intrinsic topological superconductivity are scarce. So far candidate materials include Sr$_2$RuO$_4$ \cite{Mackenzie2017}, transition-metal dichalcogenides such as T$_{\rm d}$-MoTe$_2$ \cite{Guguchia2017} and 4H$_{\rm b}$-TaS$_2$ \cite{Ribak2020}, uranium-based heavy-fermion systems \cite{Ran2019,Metz2019}, 
$\beta$-PdBi$_2$ \cite{Sakano_2015, Biswas_2016, Kolapo_2019, Li_2024}, 
and very recently, the kagome material RbV$_3$Sb$_5$ \cite{Wang2024}.
In these systems, however, different experimental methods produce inconclusive and sometimes contradictory results, so that to date no material has been convincingly shown to be an intrinsic topological superconductor \cite{vonRohr2023}.

The unconventional $i$-wave SC order implies the presence of six Majorana cones on a given PtBi$_2$ surface, each with their owntopological invariant -- a winding number equal to either $+1$ or $-1$. 
Interestingly, symmetry dictates all six have the same winding number. 
This is the signature of a quantum anomaly: on the opposite surface of a SC slab there are six Majorana modes of opposite winding number, ensuring that the sum over all topological invariants vanishes. 
We will show theoretically that the edge-state structure related to these topological nodes causes the existence of zero-energy, dispersionless Majorana modes localized at the sample hinges, which in practice may be realized by sufficiently high step edges at the surface.

\section*{ARPES characterization}

Trigonal PtBi$_2$ is a noncentrosymmetric Weyl se\-mi\-me\-tal belonging to space group $P31m$ \cite{Shipunov2020}. 
Its electronic structure hosts 12 Weyl cones, related to each other by time reversal as well as threefold rotation symmetry, which are positioned about $47\,\mathrm{meV}$ above the Fermi level \cite{Veyrat2023, Kuibarov2024, Vocaturo2024, Schimmel2024}. 
Upon cleaving, two different types of surfaces are produced, a kagome-type (KT) surface and a decorated honeycomb (DH) surface \cite{Kuibarov2024, Vocaturo2024}. At both terminations scanning tunnelling spectroscopy has established the presence of superconductivity \cite{Schimmel2024}, carried by the topological surface states of this Weyl semimetal -- the Fermi arcs -- which gap out at temperatures below $10\,\mathrm{K}$ \cite{Kuibarov2024}.

\begin{figure}
\centering
\includegraphics[width=.7\linewidth]{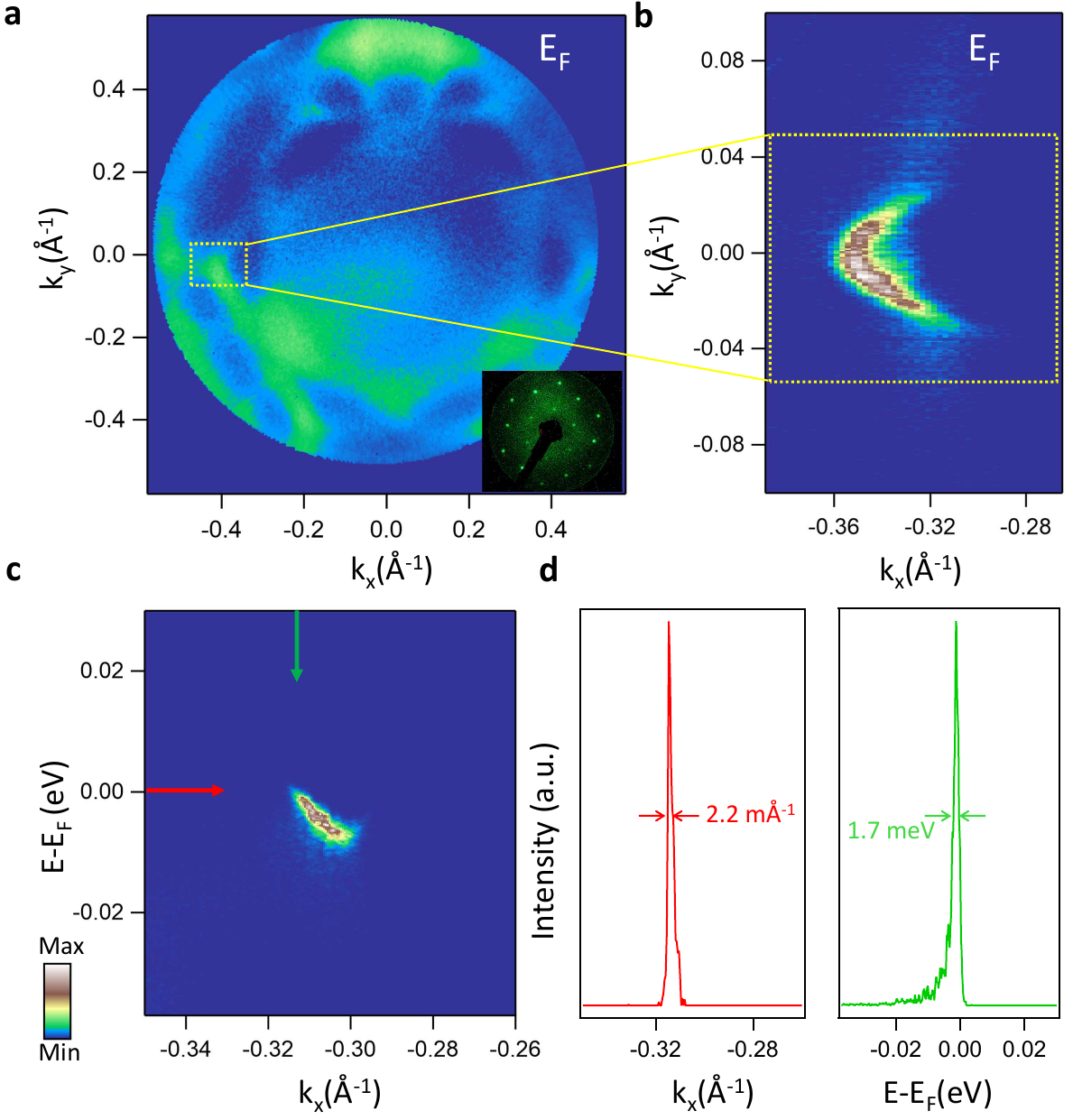}
\caption{
{\bf Progress in experimental accuracy.}
{\bf (a)} Fermi surface (FS) map observed with FeSuMa and He-I lamp $h\nu=21.2\,\mathrm{eV}$ from KT-termination. 
Collected LEED image on PtBi$_2$ single crystal is demonstrated in the inset. 
The yellow box marks the position of the arc at the FS map. 
{\bf (b)} The arc becomes well resolved in the FS observed with Laser ARPES with $h\nu=6\,\mathrm{eV}$ (KT-termination). 
{\bf (c)} Momentum-energy intensity distribution corresponding to the momentum cut through the arc (DH-termination). {\bf (d)} MDC and EDC plotted along the red and green arrows of panel (c).
}
\label{fig:1}
\end{figure}

We have carried out ARPES experiments with improved resolution to specifically study the gap function on the Fermi arcs of PtBi$_2$. 
We start by demonstrating in Fig.~\ref{fig:1} the progress in experimental accuracy in comparison with our previous study \cite{Kuibarov2024}. 
It has already been shown that the most precise measurements can be carried out with the lowest possible photon energy, which leads to the lowest kinetic energy of the photoelectrons of interest. 
However, such low kinetic energies correspond to relatively small values of the absolute momentum, which do not cover a sufficient portion of the Brillouin zone (BZ). 
Therefore, we first perform the Fermi-surface mapping using the higher photon energies available in the laboratory ($21.2\,\mathrm{eV}$ from a helium lamp) and use for this purpose a FeSuMa analyzer \cite{Borisenko2022}, which allows to record such maps with isotropic angular resolution. 
This data set together with the LEED picture is shown in Fig.~\ref{fig:1}(a). 

Although the presence of arcs along $\Gamma$--M  is visible on the map, the arcs themselves are not well pronounced at this photon energy, in agreement with our previous observation at a synchrotron \cite{Kuibarov2024}. 
After checking the quality of the surface and orientation of the sample, we use a laser source with $h\nu=6\,\mathrm{eV}$ to zoom in to the vicinity of the M point of the BZ and collect ARPES data with better momentum and energy resolution using a conventional analyzer [Fig.~\ref{fig:1}(b)]. 
With this method, the arc is seen with unprecedented clarity and in close agreement with the DFT slab calculations, e.g., in Ref.~\cite{Kuibarov2024}.  
Another advantage of using this particular photon energy is the strong enhancement of the arc intensity compared to the bulk bands. 
The same holds when considering a momentum-energy cut through the arc, as shown in Fig.~\ref{fig:1}(c).  
There are basically no other features visible except for the surface band supporting the Fermi arc. 
Instead of the parabolic dispersion usually underlying closed electron-like pockets of the Fermi contours, one directly observes a clearly asymmetric shape, just as expected from the open nature of the Fermi arcs in Weyl semimetals. 

We further reduce the dimension of the data set by extracting the momentum distribution curve (MDC) and energy distribution curve (EDC) along the red and green arrows in Fig.~\ref{fig:1}(c), respectively. 
The resulting very sharp and strong peaks are shown in Fig.~\ref{fig:1}(d). 
Both full width at half maximum (FWHM) values, namely $2.2$ {m\AA}$^{-1}$ and $1.7\,\mathrm{meV}$, are the smallest in the history of photoemission from solids for such lineshapes. 
Thus, the Fermi arcs in PtBi$_2$ appear to represent extraordinary electronic states, remarkably strongly localized in terms of energy, momentum, and space. 
The ability to detect such features with the precision shown above provides a natural opportunity to investigate the order parameter when the arcs become superconducting.

\section*{Evidence for nodes in the superconducting gap}

\begin{figure*}
\centering
\includegraphics[width=0.9\linewidth]{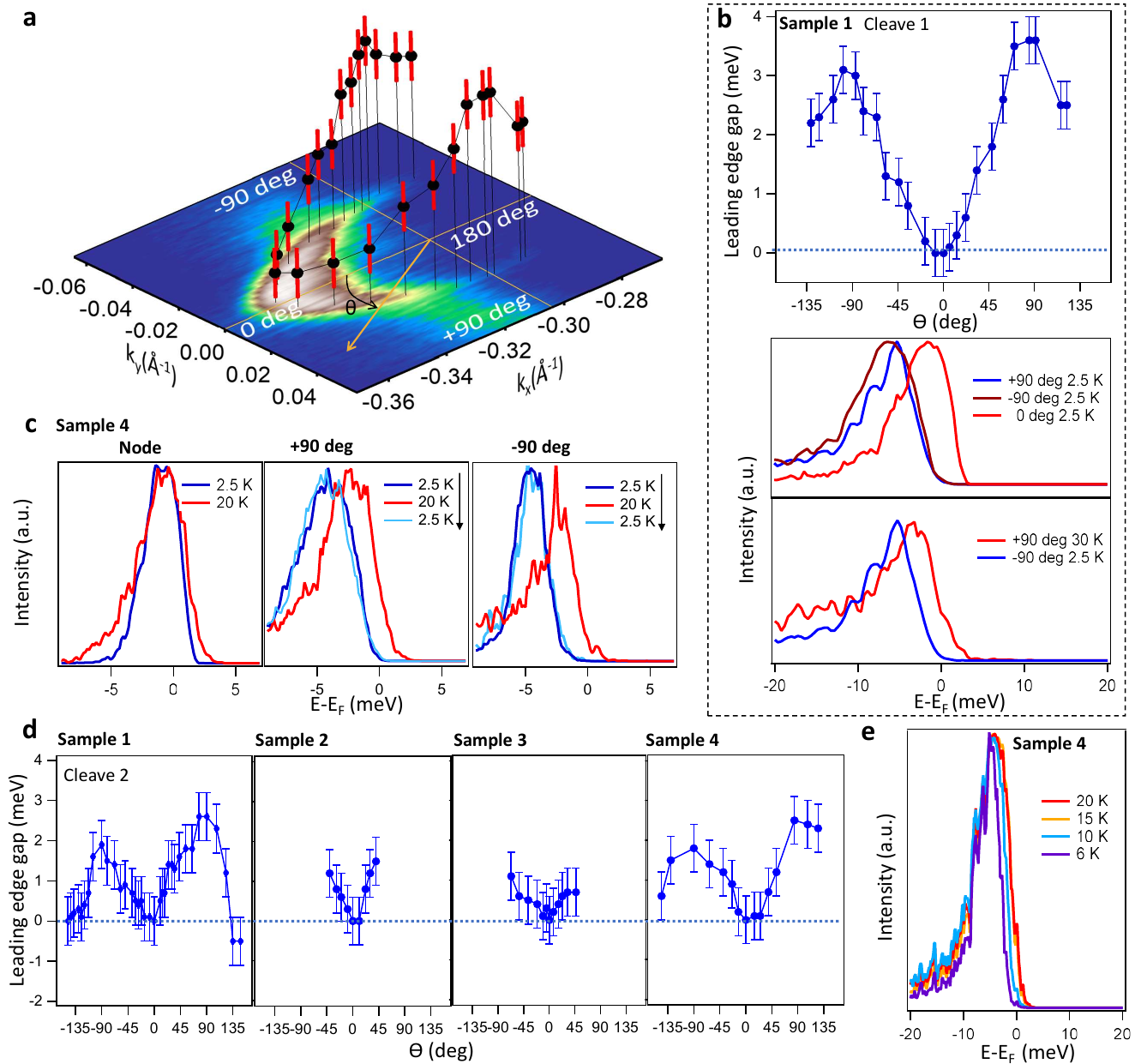}
\caption{
{\bf Anisotropic superconducting gap.}
{\bf (a)} leading edge gap across different points of the arc (KT-termination). 
{\bf (b)} Angular dependence of the gap, showing a node at $\theta=0^\circ$ and a maximum gap at $\pm 90^\circ$ for Cleave 1 of sample 1 (top panel). Middle panel shows the leading edge gap from EDCs taken along $\theta = 0^\circ$ and $\pm 90^\circ$ at $2.5\,\mathrm{K}$. 
This is equivalent to the gap observed from EDCs taken at $+90^\circ$ with $2.5\,\mathrm{K}$ and $30\,\mathrm{K}$ (bottom panel).
{\bf (c)} Energy distribution curves (EDCs) taken at the node, $+90^{\circ}$, and $-90^{\circ}$ respectively for 2.5 K and 20 K. For $\pm 90^{\circ}$, the temperature is cycled back to 2.5 K, which overlap with the initial 2.5 K EDCs.  
{\bf (d)} Angular dependence for Cleave 2 of sample 1 (left panel) and for three other PtBi$_2$ single crystals from different batches. 
{\bf (e)} Temperature dependence of the EDC corresponding to the arc exhibiting the gradual closing of the leading edge gap at higher temperature.
}
\label{fig:2}
\end{figure*}

Next, we focus on determining the leading edge gap at different points along the arc in Fig.~\ref{fig:2}. 
The first remarkable observation is that this gap is not isotropic. 
The 3D image in Fig.~\ref{fig:2}(a) shows the points of the arc where the leading edge gap is determined and how the gap changes along the arc. 
The momentum $k_y=0$ {\AA}$^{-1}$ corresponds to the $\Gamma$--M direction in the BZ. 
The gap is also plotted in Fig.~\ref{fig:2}(b), clearly showing this anisotropy. 
An immediate and rather surprising observation is that the leading edge gap appears to close when the arc crosses the $\Gamma$--M line, indicating the existence of a node. 
Due to the finite resolution, it is not possible to determine the exact behaviour of the gap function very close to this point, as is the case for the high-$T_c$ cuprates, but indeed also our temperature-dependent measurements confirm nodal behavior. In Fig.~\ref{fig:2}(c), we show EDCs taken from the node (0$^\circ$) and $\pm90^\circ$ along the arc above and below the critical temperature. As suggested by the gap function from the top panel of Fig.~\ref{fig:2}(b), the gap increases with distance from the node, resulting in a shift of the coherence peak.

The gap reaches its maximum at approximately $\theta=\pm90^\circ$ and then starts to decrease again at higher $\theta$ [Fig.~\ref{fig:2}(b)]. 
In the middle panel of Fig.~\ref{fig:2}(b), we plot the EDCs taken along the yellow lines marked as $\theta=0$ and $\pm90^\circ$ in Fig.~\ref{fig:2}(a), which correspond to these maxima. 
The coherence peak shifts by about $3.6\,\mathrm{meV}$ towards higher kinetic energy for $\theta=\pm90^\circ$ compared to $\theta=0^\circ$. This value is closely in agreement with our earlier study and other experiments, see Ref.~\cite{Kuibarov2024} and references therein. 

We also note the apparent presence of plateaus at $\theta=\pm45^\circ$, which could be an indication of the admixture of even higher orders (see e.g. \cite{Parker_2007}), however at the current accuracy we cannot rule out that this feature is a robust observation beyond the error bars.

The gap is also observed from the EDCs taken above and below the critical temperature $T_c$ along $\theta=+90^\circ$ [bottom panel in Fig.~\ref{fig:2}(b)]. To further confirm the existence of the node on the arc, we have repeated the measurements for four different PtBi$_2$ samples grown in different batches. 
All data presented in Figs.~\ref{fig:2}(a) and (b) were taken from sample 1 Cleave 1. The data presented in Fig.~\ref{fig:2}(c) was taken from sample 4.
The angular dependence of the gap for Cleave 2 of sample 1 is presented in the leftmost panel of Fig.~\ref{fig:2}(d). 
As expected, the angular dependence of the gap is quite similar to Cleave 1 [Fig.~\ref{fig:2}(b)] with its maximum of about $2.5\,\mathrm{meV}$ at $\theta=\pm90^\circ$. 
Apart form the node at $0^\circ$, the leading edge gap gradually decreases and closes again at $\pm 125^\circ$. 
As the arc blends with the bulk bands at higher $\theta$, the effect of superconductivity, which is intrinsic to the arc, starts to disappear. 
For further validation of our observation of the node at $0^\circ$, we have repeated the same measurement on three other PtBi$_2$ samples. 
Samples 2, 3, and 4 exhibit a node at the same position as sample 1, as seen in Fig.\ \ref{fig:2}(d).
Moreover, we performed temperature dependence measurements on sample 4, which shows the gradual closure of the gap with temperature [Fig.~\ref{fig:2}(e)].
The gap decreases as we increase temperature from $6\,\mathrm{K}$ to $10\,\mathrm{K}$ but remains open. 
At $15\,\mathrm{K}$, the gap seems to be closed as we do not observe any peak shift between $15\,\mathrm{K}$ and $20\,\mathrm{K}$. 
Hence, the critical temperature is within the range of $10\,\mathrm{K}$ to $15\,\mathrm{K}$, as established in earlier studies \cite{Kuibarov2024}.

It is important to note that, the extracted leading edge position from any EDC is influenced by energy-momentum resolution, position of the Fermi level, and Fermi function (i.e. temperature of the sample). These effects can be minimized by tracking the peak position or the trailing edge position of the same EDCs along the arc as they are located away from the Fermi level, at higher binding energy. 

\subsection*{Superconducting pairing symmetry}

Based on the $C_{3v}$ point group of trigonal PtBi$_2$, the possible SC states can be classified according to its irreducible representations (irreps) $A_1$, $A_2$, and $E$. 
States belonging to the trivial irrep $A_1$ are invariant (even) under all point-group operations so that symmetry thus does not impose gap nodes. 
SC states of $E$ symmetry must either break rotation symmetry (and then can have nodes on some but not all Fermi arcs) or time-reversal symmetry (and are nodeless). 
There is no experimental evidence for the SC state breaking time-reversal symmetry, nor for the simultaneous presence of fully gapped and nodal Fermi arcs, speaking against $E$ symmetry. 
This leaves SC states of $A_2$ symmetry, which indeed have symmetry-imposed gap nodes at the arc centers and the same gap profile for all arcs.
For the $A_2$ irrep, the lowest-order time-reversal-symmetric basis function of the polar angle $\phi$ in 2D momentum space is $\sin l \phi$ with $l=6$, implying $i$-wave pairing symmetry. 
Since the Fermi arc is chiral and nondegenerate, a sign change of the SC order parameter along the arc directly produces a surface Majorana cone, the hallmark of topological superconductivity. 
This is similar to the Majorana cones that are expected to occur on the surface of 3D strong topological SCs \cite{Schnyder2008} or $^3$He \cite{Salomaa1988}. 

\section*{Comparison to electronic structure calculations}

To compare the ARPES results to density functional theory (DFT) we modified the approach of Ref.~\cite{Kuibarov2024} to include nodal gap functions. 
In particular we use in our DFT Wannier model $i$-wave pairing of the form $V_0 \sin\left(6\phi\right)$, expanded for small momenta around the node. 
We restrict pairing and thus $V_{0}\ne0$ to the surface block of a semi infinite slab and solve the Bogoliubov-de-Gennes (BdG) equations in the semi-infinite slab geometry to obtain the surface Bloch spectral density $A_{\mathrm{bl}}\left(\boldsymbol{k},E\right)$.

\begin{figure*}
\begin{centering}
\includegraphics[width=\linewidth]{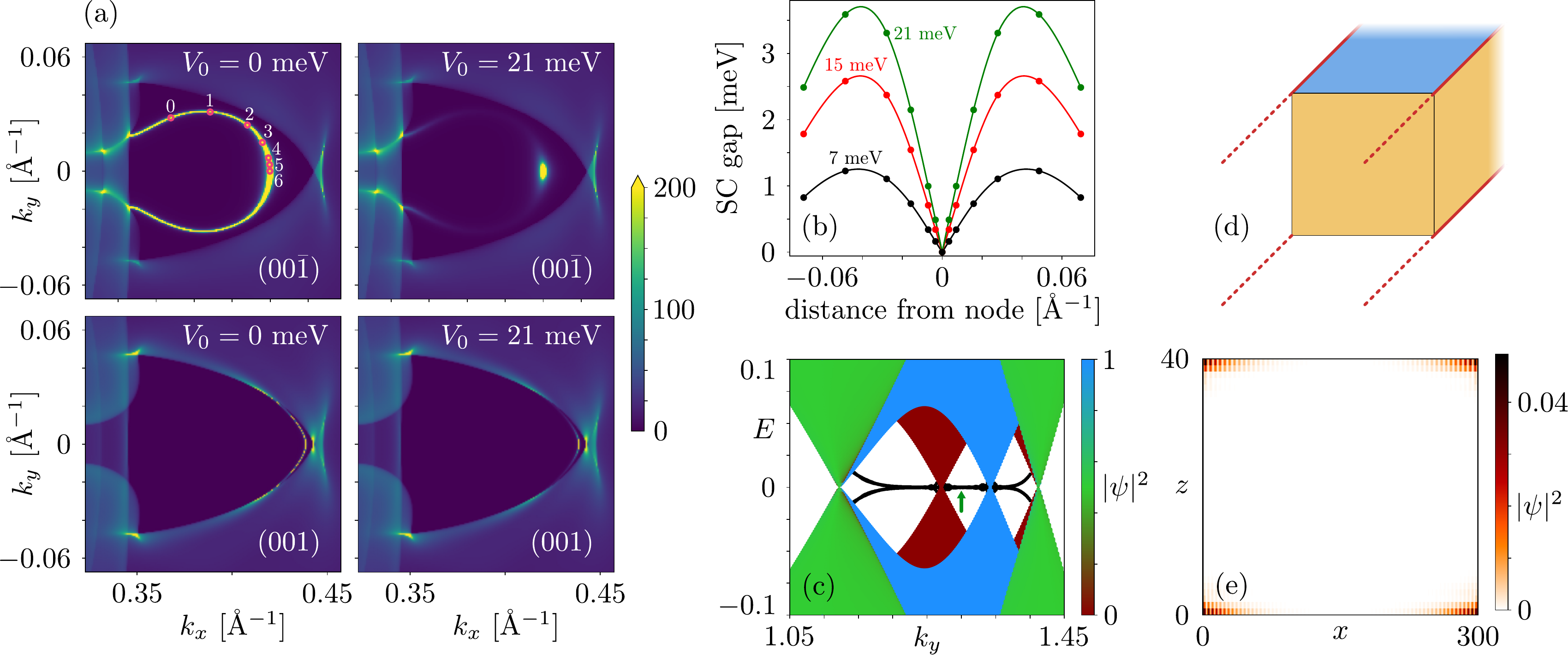}
\par\end{centering}
\caption{\label{fig:calculations} 
{\bf Calculated properties of the $i$-wave superconductor -- Majorana cones and hinge states.}
\textbf{(a)}  
Calculated spectral density at the Fermi level in a DFT-BdG Wannier model for the $\left(00\bar{1}\right)$ surface (DH termination \cite{Vocaturo2024}, top panels) and for the $(001)$ surface (bottom panels)  with superconducting $i$-wave pairing $V_{0} \sin\left(6\varphi\right)$ on the first three surface layers for coupling strengths $V_{0}=0$ (no superconductivity) and $V_0 = 21$ meV. 
The points at which the gap was determined are indicated in the normal state panel using labels 0 to 6. 
Spectral weights larger than $200$ are shown in yellow.
\textbf{(b)} The SC gap as a function of distance from the node for three coupling strengths $V_{0}=7$ (black), 15 (red), and 21 meV (green), in the case of the DH termination, as in the top panels of (a). 
Circles are calculated values, the lines are guides to the eyes. 
{\bf (c)} Electronic structure of the effective model. The color scale denotes the probability density of the states in real space. 
Bulk Weyl cones are shown in green, and top and bottom Majorana cones are shown in red and blue, respectively.  
Dispersionless zero-energy Majorana hinge modes are shown in black.  
{\bf{(d)}} Sketch of the prism geometry used in the effective model. 
The system is infinite in the $y$ direction and finite in $x$ and $z$. 
The superconducting top surface is shown in blue, and the Majorana hinge modes are shown in red. 
{\bf (e)} Probability density of the four Majorana hinge modes in the prism geometry of panel (d), computed for $k_y=1.3$, corresponding to the green arrow in panel (c). 
}
\end{figure*}

Figure \ref{fig:calculations}a shows the results for different coupling strengths $V_{0}$. 
In the normal state ($V_{0}=0$) the gap vanishes along the whole Fermi arc, while for finite $V_{0}$ the gap closes at $k_{y}=0$, the center of the Majorana-cone. 
Due to a numerical finite life time of $0.05$ meV, a remnant spectral weight is visible along the arc, especially for small $V_{0}$. 
The gap however is finite, except at the node.

The gap was determined by zooming in on the points $\boldsymbol{k}_{i}$, indicated in the normal state panel, to locate points exactly on top of the arc, followed by a scan of $A_{\mathrm{bl}}\left(\boldsymbol{k}_{i},E\right)$ to determine the quasiparticle edges.
Comparing the resulting gaps, shown in Fig.~\ref{fig:calculations}b, to experiment allows to deduce a coupling strength of $V_{0}=15$ to $20$ meV. 
The gap on the arc has two maxima that are caused by the gap vanishing at both the $\Gamma$-M node and the surface projected Weyl node position, where the arc states merge with the bulk.
Even when we include in the calculations only the lowest $i$-wave harmonic, the position of the gap maxima at 0.04 \AA$^{-1}$ is close to the experimental value 0.06 $\pm$ 0.01 \AA$^{-1}$.  

\section*{Step-edge Majorana modes}

To further explore the implications of the presence of the Majorana cones theoretically, we use a simplified effective tight-binding model for PtBi$_2$ that captures both the lattice symmetries and the relative positions of the Weyl cones in the normal state \cite{Vocaturo2024}. 
The model disregards topologically trivial bands crossing the Fermi level while ensuring a significant separation between Weyl cones in momentum space. 

Without surface superconductivity, the projections of the 12 Weyl cones are connected pairwise by surface Fermi arcs. 
Upon including the surface pairing terms with $A_2$ symmetry in a slab geometry (infinite along the $x$ and $y$ directions, finite along $z$), the Fermi arcs on both the top and the bottom surfaces are gapped out by the pairing, leading to the formation of gapless Majorana cones along the $\Gamma$--M direction of the slab BZ. 
As in the DFT-BdG description, each Fermi arc produces one surface Majorana cone, such that there exist six cones on the top surface and six on the bottom surface. 
In the presence of both time-reversal and particle-hole symmetries, corresponding to class DIII in the Altland-Zirnbauer classification \cite{Altland1997}, each individual Majorana cone is topologically protected, being characterized by a nonzero winding number, taking the value $\pm 1$ \cite{Wen2002}. 
Note that this scenario is distinct from that of gapped strong topological superconductors in class DIII, in which the surface hosts an integer number of Majorana cones due to 3D bulk topology.
Here, instead, the bulk remains metallic and does not superconduct, whereas the surface realizes a 2D gapless topological phase \cite{Beri2010}.
Interestingly, since the surface Majorana cones on a given surface are related to each other by threefold rotation and/or time-reversal symmetry, they all have the same sign of the winding number \cite{Beri2010}, which in our case is $-1$ for the top surface and $+1$ for the bottom one. 
Thus, each surface of the system forms a so-called {\it anomalous} topological superconductor, one in which the sum of Majorana-cone winding numbers does not cancel, and therefore one which is impossible to realize in a purely 2D system.
In contrast, in the gapless topological phases of purely 2D superconductors, the winding numbers of Majorana cones must vanish, and the number of cones must be a multiple of four, as shown in Ref.~\cite{Beri2010}.

Breaking time-reversal symmetry removes the topological protection of the Majorana cones, which then acquire a gap. 
Interestingly, this implies that such a weak magnetic field enhances the surface SC gap close to the node, while at the same time reducing it for other momenta along the arc, a prediction which may be tested experimentally. 
When time-reversal symmetry is preserved, however, the nonzero winding number of the surface Majorana cones necessarily implies the existence of zero-energy Majorana modes \cite{Sato_2011} localized at the {\it boundaries of the surface}, i.e., at the hinges of the 3D system. 
This symmetry-based observation is confirmed in Fig.~\ref{fig:calculations}(c-e) by a calculation on an infinite prism geometry [infinite along the $y$ direction and finite along $x$ and $z$, as shown in Fig.~\ref{fig:calculations}(d)].
The prism bandstructure [Fig.~\ref{fig:calculations}(c)] contains the projection of two Majorana cones located on the top surface, which overlap in $k_y$ in the prism BZ (shown in blue, total winding number $+2$), and two Majorana cones located on the bottom surface (red, winding number $-2$).
At $k_y$ values between them [green arrow in Fig.~\ref{fig:calculations}(c)], there appear four degenerate zero modes, shown in black, which are localized at the hinges of the infinite prism [see Fig.~\ref{fig:calculations}(e)].
Outside of this momentum range, the hinge modes are no longer topologically protected, such that they can hybridize and split away from zero energy.
Similar to the surface Majorana cones themselves, the hinge modes will move away from the Fermi level under a Zeeman field, providing another signature of topological SC that is experimentally accessible in local measurements at hinges or large step edges of PtBi$_2$.

Our observation of nodal superconductivity in the Fermi-arcs of PtBi$_2$ sparks the intriguing question as to what drives this unconventional $i$-wave pairing. 
While in cuprates the mechanism for high-temperature $d$-wave superconductivity remains under debate, there is consensus that the presence of strong electron-electron interactions stabilizes the nodal $d$-wave pairing channel over the nodeless $s$-wave one. 
In PtBi$_2$ the electronic states are highly delocalized in nature and strong electronic correlations are not expected, and are indeed as yet without experimental indication. 
On the other hand the topological character of the superconducting Fermi-arcs sets PtBi$_2$ apart from any other known superconductor to date. 
The mechanism by which an $i$-wave superconductor emerges from pairing of these topological states is yet to be established.

Finally, we note that the coexistence of gapless Majorana cones with a metallic bulk impedes the applicability of PtBi$_2$ towards quantum computation, at least in its current form.
This can potentially be mitigated by fabrication of ultra-thin samples: when the thickness of the material becomes small enough, the contribution of unwanted gapless bulk modes will be reduced, or even eliminated. 
Another potential manipulation may involve breaking time-reversal symmetry in order to gap out the surface Majorana cones in such a way as to leave behind either chiral Majorana edge modes, or zero-dimensional Majorana bound states localized at the corners of the material.
Both types of gapless mode have been proposed as a potential avenue towards topological quantum computation \cite{Lian_2018, Pan_2022}.
Alternatively, one might also envisage controlling the phase difference between the top- and bottom-surface superconductors, realizing a planar Josephson junction that provides an avenue towards quantum computation \cite{arxiv.2504.20031}.

\section*{Acknowledgements}
We thank Pavan Hosur, Fabian Jakubczyk, Oleg Janson, Riccardo Vocaturo and Harald Waje for fruitful discussions and Ulrike Nitzsche for technical assistance. This work was supported by the Deutsche Forschungsgemeinschaft (DFG, German Research Foundation) under Germany's Excellence Strategy through the W\"{u}rzburg--Dresden Cluster of Excellence ``Complexity and Topology in Quantum Matter'' -- ct.qmat (EXC 2147, project id 390858490) and within the Collaborative Research Center ``Correlated Magnetism: From Frustration to Topology'' (SFB 1143, project id 247310070). This work was also supported by BMBF funding through project 01DK240008  (GU-QuMat).

\clearpage	

%\let\oldaddcontentsline\addcontentsline% Store \addcontentsline
%\renewcommand{\addcontentsline}[3]{}% Make \addcontentsline a no-op
%\bibliography{ptbi2sc}
%\let\addcontentsline\oldaddcontentsline% Restore \addcontentsline

%apsrev4-2.bst 2019-01-14 (MD) hand-edited version of apsrev4-1.bst
%Control: key (0)
%Control: author (8) initials jnrlst
%Control: editor formatted (1) identically to author
%Control: production of article title (0) allowed
%Control: page (0) single
%Control: year (1) truncated
%Control: production of eprint (0) enabled
%

\end{document}